\documentclass{article}
\usepackage{fullpage}
\usepackage[american]{babel}
\usepackage{amssymb}
\usepackage{algorithmic}
\usepackage{algorithm}
\usepackage{multirow}
\usepackage{graphicx}

\newcommand{\eqn}[1]{(\ref{#1})}
\newcommand{\beql}[1]{\begin{equation}\label{#1}}
\newcommand{\eeq}{\end{equation}}
\newtheorem{theo}{Theorem}
\newtheorem{defi}{Definiton}
\newtheorem{asm}{Assumption}

\title{An Efficient Dynamic Programming Algorithm for the Generalized LCS Problem with Multiple Substring Exclusion Constrains}
\author{Lei Wang, Xiaodong Wang, Yingjie Wu, and Daxin Zhu}

\begin{document}
\maketitle

\begin{abstract}
In this paper, we consider a generalized longest common subsequence problem with multiple substring exclusion constrains. For the two input sequences $X$ and $Y$ of lengths $n$ and $m$, and a set of $d$ constrains $P=\{P_1,\cdots,P_d\}$ of total length $r$, the problem is to find a common subsequence $Z$ of $X$ and $Y$ excluding each of constrain string in $P$ as a substring and the length of $Z$ is maximized. The problem was declared to be NP-hard\cite{1}, but we finally found that this is not true.
A new dynamic programming solution for this problem is presented in this paper. The correctness of the new algorithm is proved. The time complexity of our algorithm is $O(nmr)$.
\end{abstract}

\section{Introduction}
In this paper, we consider a generalized longest common subsequence problem with multiple substring exclusion constrains. The longest common subsequence (LCS) problem is a well-known measurement for computing the similarity of two strings. It can be widely applied in diverse areas, such as file comparison, pattern matching and computational biology\cite{3,4,8,9}.

Given two sequences $X$ and $Y$, the longest common subsequence (LCS) problem is to find a subsequence of $X$ and $Y$ whose length is the longest among all common subsequences of the two given sequences.

For some biological applications some constraints must be applied to the LCS problem. These kinds of variant of the LCS problem are called the constrained LCS (CLCS) problem. Recently, Chen and Chao\cite{1} proposed the more generalized forms of the CLCS problem, the generalized constrained longest common subsequence (GC-LCS) problem.
For the two input sequences $X$ and $Y$ of lengths $n$ and $m$,respectively, and a constraint string $P$ of length $r$, the GC-LCS problem is a set of four problems which are to find the LCS of $X$ and $Y$ including/excluding $P$ as a subsequence/substring, respectively. The four generalized constrained LCS can be summarized in Table 1.

\begin{table}[ht]
\begin{center}
\caption{The GC-LCS problems}
\begin{tabular}{|l|l|l|}
\hline\hline
Problem & Input & Output\\\hline
\multirow{2}{*}{SEQ-IC-LCS} & \multirow{2}{*}{$X$,$Y$, and $P$} & The longest common subsequence of $X$ and $Y$ \\
 &  & including $P$ as a subsequence \\\hline
\multirow{2}{*}{STR-IC-LCS} & \multirow{2}{*}{$X$,$Y$, and $P$} & The longest common subsequence of $X$ and $Y$ \\
 &  & including $P$ as a substring \\\hline
\multirow{2}{*}{SEQ-EC-LCS} & \multirow{2}{*}{$X$,$Y$, and $P$} & The longest common subsequence of $X$ and $Y$ \\
 &  & excluding $P$ as a subsequence\\\hline
\multirow{2}{*}{STR-EC-LCS} & \multirow{2}{*}{$X$,$Y$, and $P$} & The longest common subsequence of $X$ and $Y$ \\
 &  & excluding $P$ as a substring\\\hline
\end{tabular}
\end{center}
\end{table}

The four GC-LCS problems can be generalized further to the cases of multiple constrains. In these generalized cases, the single constrained pattern $P$ will be generalized to a set of $d$ constrains $P=\{P_1,\cdots,P_d\}$ of total length $r$, as shown in Table 2.

\begin{table}[ht]
\begin{center}
\caption{The Multiple-GC-LCS problems}
\begin{tabular}{|l|l|l|}
\hline\hline
Problem & Input & Output\\\hline
\multirow{2}{*}{M-SEQ-IC-LCS} & $X$,$Y$, and a set of constrains & The longest common subsequence of $X$ and $Y$ \\
 &  $P=\{P_1,\cdots,P_d\}$  & including each of constrain $P_i\in P$ as a subsequence \\\hline
\multirow{2}{*}{M-STR-IC-LCS} & $X$,$Y$, and a set of constrains & The longest common subsequence of $X$ and $Y$ \\
 &  $P=\{P_1,\cdots,P_d\}$  & including each of constrain $P_i\in P$ as a substring \\\hline
\multirow{2}{*}{M-SEQ-EC-LCS} & $X$,$Y$, and a set of constrains & The longest common subsequence of $X$ and $Y$ \\
 &  $P=\{P_1,\cdots,P_d\}$  & excluding each of constrain $P_i\in P$ as a subsequence\\\hline
\multirow{2}{*}{M-STR-EC-LCS} & $X$,$Y$, and a set of constrains & The longest common subsequence of $X$ and $Y$ \\
 &  $P=\{P_1,\cdots,P_d\}$  & excluding each of constrain $P_i\in P$ as a substring\\\hline
\end{tabular}
\end{center}
\end{table}

The Multiple-GC-LCS problem M-SEQ-IC-LCS has been proved to be NP-hard in \cite{5}.
The Multiple-GC-LCS problem M-SEQ-EC-LCS has also been proved to be NP-hard in \cite{6,10}.
In addition, The Multiple-GC-LCS problems M-STR-IC-LCS and M-STR-EC-LCS were also declared to be NP-hard in \cite{1}, but without strict proofs.
The exponential-time algorithms for solving these two problems were also presented in \cite{1}.

We will discuss the multiple STR-EC-LCS problem M-STR-EC-LCS in this paper. A cubic time algorithm is presented for the  M-STR-EC-LCS problem and disproves that this problem is NP-hard.

The organization of the paper is as follows.

In the following 4 sections we describe our presented dynamic programming algorithm for the M-STR-EC-LCS problem.

In section 2 the preliminary knowledge for presenting our algorithm for the M-STR-EC-LCS problem is discussed.
In section 3 we give a new dynamic programming solution for the M-STR-EC-LCS problem with time complexity $O(nmr)$, where $n$ and $m$ are the lengths of the two given input strings, and $r$ is the total length of $d$ constrain strings.
In section 4 we discuss the issues to implement the algorithm efficiently.
Some concluding remarks are in section 5.

\section{Preliminaries}
A sequence is a string of characters over an alphabet $\sum$. A subsequence of a sequence $X$ is obtained by deleting zero or more characters from $X$ (not necessarily contiguous). A substring of a sequence $X$ is a subsequence of successive characters within $X$.

For a given sequence $X=x_1x_2\cdots x_n$ of length $n$, the $i$th character of $X$ is denoted as $x_i \in \sum$ for any $i=1,\cdots,n$. A substring of $X$ from position $i$ to $j$ can be denoted as $X[i:j]=x_ix_{i+1}\cdots x_j$. If $i\neq 1$ or $j\neq n$, then the substring $X[i:j]=x_ix_{i+1}\cdots x_j$ is called a proper substring of $X$. A substring $X[i:j]=x_ix_{i+1}\cdots x_j$ is called a prefix or a suffix of $X$ if $i=1$ or $j=n$, respectively.

For the two input sequences $X=x_1x_2\cdots x_n$ and $Y=y_1y_2\cdots y_m$ of lengths $n$ and $m$, respectively, and a set of $d$ constrains $P=\{P_1,\cdots,P_d\}$ of total length $r$, the multiple STR-EC-LCS problem M-STR-EC-LCS is to find an LCS of $X$ and $Y$ excluding each of constrain $P_i\in P$ as a substring.

The most important difference between the problems STR-EC-LCS and M-STR-EC-LCS is the number of constrains. For ease of discussion, we will make the following two assumptions on the constrain set $P$.

\begin{asm}
There are no duplicated strings in the constrain set $P$.
\end{asm}

\begin{asm}
No string in the constrain set $P$ is a proper substring of any other string in $P$.
\end{asm}

Keyword tree\cite{2,7} is a main data structure in our dynamic programming algorithm to process the constrain set $P$ of the M-STR-EC-LCS problem.

\begin{defi}\label{df1}
The Keyword tree for set $P$ is a rooted directed tree $T$ satisfying 3 conditions: 1. each edge is labeled with exactly one character; 2. any two edges out of the same node have distinct labels; and 3. every string $P_i$ in $P$ maps to some node $v$ of $T$ such that the characters on the path from the root of $T$ to $v$ exactly spell out $P_i$, and every leaf of $T$ is mapped to by some string in $P$.
\end{defi}

\begin{figure}
\centering
\includegraphics[width=15.5cm,height=5.5cm]{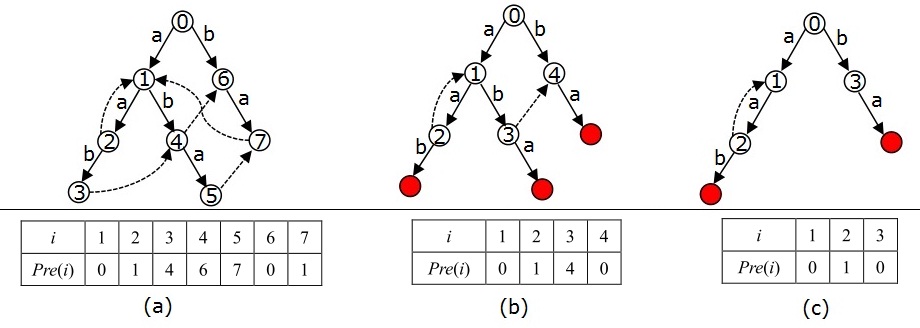}
\caption{Keyword Trees}
\end{figure}

For example, Figure 1(a) shows the keyword tree $T$ for the constrain set $P=\{aab,aba,ba\}$, where $q=3,r=8$.
Clearly, every node in the keyword tree corresponds to a prefix of one of the strings in set $P$, and every prefix of a string $P_i$ in $P$ maps to a distinct node in the keyword tree $T$.
The keyword tree for set $P$ of total length $r$ of all strings can be easily constructed in $O(r)$ time for a constant alphabet size.
Because no two edges out of any node of $T$ are labeled with the same character, the keyword tree $T$ can be used to search for all occurrences in a text $X$ of strings from $P$.

The failure functions in the Knuth-Morris-Pratt algorithm for solving the string matching problem can be generalized to the case of keyword tree to speedup the exact string matching of multiple patterns as follows.

In order to identify the states of the nodes of $T$, we assign numbers $0,1,\cdots,t-1$ to all $t$ nodes of $T$ in their preorder numbering. Then, each node will be assigned an integer $i,0\leq i<t$, as shown in Fig.1. In the following, we also use the node number as its state number of the node in $T$.

For each node numbered $i$ of a keyword tree $T$, the concatenation of characters on the path from the root to the node $i$ spells out a string denoted as $L(i)$. The string $L(i)$ is also called the label of node $i$ in the keyword tree $T$.
For any node $i$ of $T$, define $lp(i)$ to be the length of the longest proper suffix of string $L(i)$ that is a prefix of some string in $T$.

It can be verified readily that for each node $i$ of $T$, if $A$ is an $lp(i)$-length suffix of string $L(i)$, then there must be a unique node $pre(i)$ in $T$ such that $L(pre(i))=A$. If $lp(i)=0$ then $pre(i)=0$ is the root of $T$.

\begin{defi}\label{df2}
The ordered pair $(i,pre(i))$ is called a failure link.
\end{defi}

The failure link is a direct generalization of the failure functions in the KMP algorithm.
For example, in Figure 1(a), failure links are shown as pointers from every node $i$ to node $pre(i)$ where $lp(i)>0$. The other failure links point to the root and are not shown.

The failure links of $T$ define actually a failure function $pre$ for the constrain set $P$.

For example, for the nodes $i=1,2,3,4,5,6,7$ in Fig.1, the corresponding values of failure function are $pre(i)=0,1,4,6,7,0,1$, as shown in Fig.1.

The failure function $pre$ is used to speedup the search for all occurrences in a text $X$ of strings from $P$.
As stated in \cite{7}, the failure function $pre$ can be computed in $O(r)$ time.

In the keyword tree application in our dynamic programming algorithm, a function $\sigma$ will be mentioned frequently.
For a string $S$ and a given keyword tree $T$, if the label $L(i)$ of a node numbered $i$ is also a suffix of $S$, then the node $i$ is called a suffix node of $S$ in $T$.

\begin{defi}\label{df3}
For any string $S$ and a given keyword tree $T$, the unique suffix node of $S$ in $T$ with maximum depth is denoted as $\sigma(S)$.
That is:
\beql{eq21}
|L(\sigma(S))|=\max_{0\leq i<t}\{|L(i)|| L(i) \verb" is a suffix of " S\}
\eeq
\end{defi}

For example, if $S=aabaaabb$, then in the keyword tree $T$ of Fig.1, the node 6 is the only suffix node of $S$ in $T$, therefore $\sigma(S)=6$.

In our keyword tree application, we are only interested in the nonleaf nodes of the tree. So, we can renumber the nodes of the tree only for nonleaf nodes, omitting the leaf nodes of the tree, as shown in Fig.1(b). After renumbering, the failure function of the tree will also be changed accordingly.

If a string $P_i$ in the constrain set $P$ is a proper substring of another string $P_j$ in $P$, then an LCS of $X$ and $Y$ excluding $P_i$ must also exclude $P_j$. For this reason, the constrain string $P_j$ can be removed from constrain set $P$ without changing the solution of the problem.
For example, the string $ba$ is a proper substring of the string $aba$ in the keyword tree of Fig.1(a). Therefore, the string $aba$ can be removed from the keyword tree, as shown in Fig.1(c). We will show shortly how to remove these redundant strings from constrain set $P$ in $O(r)$ time.
In the following sections, discussions are based on the Assumption 1 and 2 on the constrain set $P$. The number of nonleaf nodes of the keyword tree for the constrain set $P$ is denoted as $s$. In the worst case $s=r-d$. The root of the keyword tree is numbered 0, and the other nonleaf nodes are numbered $1,2,\cdots,s-1$ in their preorder numbering. For example, in Fig.1(c), there are $s=3$ nonleaf nodes in $T$. The labels for the three nonleaf nodes are $L(0)=\emptyset, L(1)=a, L(2)=aa$ and $L(3)=b$ respectively.

The symbol $\oplus$ is also used to denote the string concatenation. For example, if $S_1=aaa$ and $S_2=bbb$, then it is readily seen that  $S_1\oplus S_2=aaabbb$.

\section{Our Main Result: A Dynamic Programming Algorithm}
In the following discussions, we will call 'a sequence excluding each of constrain string in $P$ as a substring' a sequence excluding $P$ for short.
\begin{defi}\label{df4}
Let $Z(i,j,k)$ denote the set of all LCSs of $X[1:i]$ and $Y[1:j]$ excluding $P$ and $\sigma(z)=k$ for each $z\in Z(i,j,k)$, where $1\leq i\leq n, 1\leq j\leq m$, and $0\leq k<s$.
The length of an LCS in $Z(i,j,k)$ is denoted as $f(i,j,k)$.
\end{defi}

If we can compute $f(i,j,k)$ for any $1\leq i\leq n, 1\leq j\leq m$, and $0\leq k<s$ efficiently, then the length of an LCS of $X$ and $Y$ excluding $P$ must be $\max\limits_{0\leq k<s}\left\{f(n,m,k)\right\}$.

By using the keyword tree data structure described in the last section, we can give a recursive formula for computing $f(i,j,k)$ by the following Theorem.

\begin{theo}\label{th1}
For the two input sequences $X=x_1x_2\cdots x_n$ and $Y=y_1y_2\cdots y_m$ of lengths $n$ and $m$, respectively, and a set of $d$ constrains $P=\{P_1,\cdots,P_d\}$ of total length $r$, let $Z(i,j,k)$ and $f(i,j,k)$ be defined as Definition \ref{df4}.
Suppose a keyword tree $T$ for the constrain set $P$ have been built, and the $s$ nonleaf nodes of $T$ are numbered in their preorder numbering.
The label of the node numbered $k(0\leq k<s)$ is denoted as $L(k)$.
Then, for any $1\leq i\leq n, 1\leq j\leq m$, and $0\leq k<s$, $f(i,j,k)$ can be computed by the following recursive formula \eqn{eq31}.

\beql{eq31}
f(i,j,k)=\left\{\begin{array}{ll}
\max\left\{ f(i-1,j,k),f(i,j-1,k) \right\} & \texttt{if } x_i\neq y_j,\\
\max\left\{
f(i-1,j-1,k),1+\max\limits_{0\leq q<s}\left\{f(i-1,j-1,q)|\sigma(L(q)\oplus x_i)=k\right\}
\right\} & \texttt{if } x_i= y_j.
\end{array} \right.
\eeq

The boundary conditions of this recursive formula are $f(i,0,k) = f(0,j,k) = 0$ for any $0\leq i\leq n, 0\leq j\leq m$, and $0\leq k \leq s$.
\end{theo}

\noindent{\bf Proof.}

For any $1\leq i\leq n, 1\leq j\leq m$, and $0\leq k<s$, suppose $f(i,j,k)=t$ and $z=z_1,\cdots, z_t\in Z(i,j,k)$.

First of all, we notice that for each pair $(i',j'), 1\leq i'\leq n, 1\leq j'\leq m$,such that $i'\leq i$ and $j'\leq j$, we have $f(i',j',k) \leq f(i,j,k)$, since a common subsequence $z$ of $X[1:i']$ and $Y[1:j']$ excluding $P$ and $\sigma(z)=k$ is also a common subsequence of $X[1:i]$ and $Y[1:j]$ excluding $P$ and $\sigma(z)=k$.

(1) In the case of $x_i\neq y_j$, we have $x_i\neq z_t$ or $y_j\neq z_t$.

(1.1)If $x_i\neq z_t$, then $z=z_1,\cdots, z_t$ is a common subsequence of $X[1:i-1]$ and $Y[1:j]$ excluding $P$ and $\sigma(z_1,\cdots, z_t)=k$, and so $f(i-1,j,k) \geq t$. On the other hand, $f(i-1,j,k)\leq f(i,j,k) = t$. Therefore, in this case we have $f(i,j,k) = f(i-1,j,k)$.

(1.2)If $y_j\neq z_t$, then we can prove similarly that in this case, $f(i,j,k) = f(i,j-1,k)$.

Combining the two subcases we conclude that in the case of $x_i\neq y_j$, we have $$f(i,j,k)=\max\left\{ f(i-1,j,k),f(i,j-1,k) \right\}$$.

(2) In the case of $x_i=y_j$, there are also two cases to be distinguished.

(2.1)If $x_i=y_j\neq z_t$,  then $z=z_1,\cdots, z_t$ is also a common subsequence of $X[1:i-1]$ and $Y[1:j-1]$ excluding $P$ and $\sigma(z_1,\cdots, z_t)=k$, and so $f(i-1,j-1,k) \geq t$. On the other hand, $f(i-1,j-1,k)\leq f(i,j,k) = t$. Therefore, in this case we have $f(i,j,k) = f(i-1,j-1,k)$.

(2.2)If $x_i=y_j=z_t$, then $f(i,j,k) = t>0$ and $z=z_1,\cdots, z_t$ is an LCS of $X[1:i]$ and $Y[1:j]$ excluding $P$ and $\sigma(z_1,\cdots, z_t)=k$, and thus $z_1,\cdots, z_{t-1}$ is a common subsequence of $X[1:i-1]$ and $Y[1:j-1]$ excluding $P$.

Let $\sigma(z_1,\cdots, z_{t-1})=q$ and $f(i-1,j-1,q)=h$.
Then $z_1,\cdots, z_{t-1}$ is a common subsequence of $X[1:i-1]$ and $Y[1:j-1]$ excluding $P$ and $\sigma(z_1,\cdots, z_{t-1})=q$.
Therefore, we have

\beql{eq32}
f(i-1,j-1,q)=h\geq t-1.
\eeq

Let $v=v_1,\cdots, v_h\in Z(i-1,j-1,q)$ is an LCS of $X[1:i-1]$ and $Y[1:j-1]$ excluding $P$ and $\sigma(v_1,\cdots, v_h)=q$. Then
$\sigma((v_1,\cdots, v_h)\oplus x_i)=\sigma(L(q)\oplus x_i)=k$, and thus $(v_1,\cdots, v_h)\oplus x_i$ is a common subsequence of $X[1:i]$ and $Y[1:j]$ excluding $P$ and $\sigma((v_1,\cdots, v_h)\oplus x_i)=k$.

Therefore,
\beql{eq33}
f(i,j,k)=t\geq h+1.
\eeq

Combining \eqn{eq32} and \eqn{eq33} we have $h=t-1$. Therefore, $z_1,\cdots, z_{t-1}$ is an LCS of $X[1:i-1]$ and $Y[1:j-1]$ excluding $P$ and $\sigma(z_1,\cdots, z_{t-1})=q$.

In other words,
\beql{eq34}
f(i,j,k)\leq 1+\max\limits_{0\leq q<s}\left\{f(i-1,j-1,q)|\sigma(L(q)\oplus x_i)=k\right\}
\eeq

On the other hand, for any $0\leq q<s$, if $f(i-1,j-1,q)=h$ and $\sigma(L(q)\oplus x_i)=k$, then for any $v=v_1,\cdots, v_h\in Z(i-1,j-1,q)$, $v\oplus x_i$ is a common subsequence of $X[1:i]$ and $Y[1:j]$ and $\sigma(v\oplus x_i)=k$. Since $v$ excludes $P$ and $\sigma(v\oplus x_i)=k<s$, $v\oplus x_i$ is a common subsequence of $X[1:i]$ and $Y[1:j]$ excluding $P$. Furthermore, $v\oplus x_i$ is a common subsequence of $X[1:i]$ and $Y[1:j]$ excluding $P$ and $\sigma(v\oplus x_i)=k$. Therefore, $f(i,j,k)=t\geq 1+h=1+f(i-1,j-1,q)$, and so we conclude that,
\beql{eq35}
f(i,j,k)\geq 1+\max\limits_{0\leq q<s}\left\{f(i-1,j-1,q)|\sigma(L(q)\oplus x_i)=k\right\}
\eeq

Combining \eqn{eq34} and \eqn{eq35} we have, in this case,
\beql{eq36}
f(i,j,k)= 1+\max\limits_{0\leq q<s}\left\{f(i-1,j-1,q)|\sigma(L(q)\oplus x_i)=k\right\}
\eeq

Combining the two subcases in the case of $x_i=y_j$, we conclude that the recursive formula \eqn{eq31} is correct for the case $x_i=y_j$.

The proof is complete.
\hfill $\blacksquare$

\section{The Implementation of the Algorithm}
According to Theorem \ref{th1}, our algorithm for computing $f(i,j,k)$ is a standard 2-dimensional dynamic programming algorithm. By the recursive formula \eqn{eq31}, the dynamic programming algorithm for computing $f(i,j,k)$ can be implemented as the following Algorithm 1.

In Algorithm 1, $s$ is the number of nonleaf nodes of the keyword tree $T$ for set $P$. The root of the keyword tree is numbered 0, and the other nonleaf nodes are numbered $1,2,\cdots,s-1$ in their preorder numbering. $L(t)$ is the label of node numbered $t$ in the keyword tree $T$.

\begin{algorithm}
\caption{M-STR-EC-LCS}
{\bf Input:} Strings $X=x_1\cdots x_n$, $Y=y_1\cdots y_m$ of lengths $n$ and $m$, respectively, and a set of $d$ constrains $P=\{P_1,\cdots,P_d\}$ of total length $r$\\
{\bf Output:} The length of an LCS of $X$ and $Y$ excluding $P$
\begin{algorithmic}[1]
\STATE Build a keyword tree $T$ for $P$
\FORALL{$i,j,k$ , $0\leq i\leq n, 0\leq j\leq m$, and $0\leq k \leq s$}
\STATE $f(i,0,k) \leftarrow 0, f(0,j,k) \leftarrow 0$ \{boundary condition\}
\ENDFOR
\FOR{$i=1$ to $n$}
\FOR{$j=1$ to $m$}
\FOR{$k=0$ to $s$}
\IF {$x_i\neq y_j$}
\STATE $f(i,j,k) \leftarrow \max\{f(i-1,j,k),f(i,j-1,k)\}$
\ELSE
\STATE $u \leftarrow \max\limits_{0\leq t<s}\left\{f(i-1,j-1,t)|\sigma(L(t)\oplus x_i)=k\right\}$
\STATE $f(i,j,k) \leftarrow \max\{f(i-1,j-1,k),1+u\}$
\ENDIF
\ENDFOR
\ENDFOR
\ENDFOR
\RETURN $\max\limits_{0\leq t<s}\{f(n,m,t)\}$
\end{algorithmic}
\end{algorithm}

To implement our algorithm efficiently, the most important thing is to compte $\sigma(L(k)\oplus x_i)$ for each $0\leq k<s$ and $x_i, 1\leq i\leq n$, in line 11 efficiently.

It is obvious that $\sigma(L(k)\oplus x_i)=g$ if there is an edge $(k,g)$ out of the node $k$ labeled $x_i$. It will be more complex to compute $\sigma(L(k)\oplus x_i)$ if there is no edge out of the node $k$ labeled $x_i$. In this case the matched node label has to be changed to the longest proper suffix of $L(k)$ that is a prefix of some string in $T$ and the corresponding node $h$ has an out edge $(h,g)$ labeled $x_i$.
Therefore, in this case, $\sigma(L(k)\oplus x_i)=g$.

\begin{algorithm}
\caption{$\sigma(k,ch)$}
{\bf Input:} Integer $k$ and character $ch$\\
{\bf Output:} $\sigma(L(k)\oplus ch)$
\begin{algorithmic}[1]
\WHILE{$k\geq 0$}
\IF{there is an edge $(k,h)$ labeled $ch$ out of the node $k$ of $T$}
\RETURN $h$
\ELSE
\STATE $k \leftarrow pre(k)$\\
\ENDIF
\ENDWHILE
\RETURN 0
\end{algorithmic}
\end{algorithm}

This computation is very similar to the search algorithm in the keyword tree $T$ for the multiple string matching problem\cite{2,7}.

With pre-computed prefix function $pre$, the function $\sigma(L(k)\oplus ch)$ for each character $ch\in\sum$ and $1\leq k\leq s$ can be described as follows.

Then, we can compute an index $t^*$ such that $$f(i-1,j-1,t^*)=\max\limits_{0\leq t<s}\left\{f(i-1,j-1,t)|\sigma(L(t)\oplus x_i)=k\right\}$$ in line 11 of Algorithm 1 by the following Algorithm 3.

\begin{algorithm}
\caption{$\max\sigma(i,j,k)$}
{\bf Input:} Integers $i,j,k$\\
{\bf Output:} An index $t^*$ such that $f(i-1,j-1,t^*)=\max\limits_{0\leq t<s}\left\{f(i-1,j-1,t)|\sigma(L(t)\oplus x_i)=k\right\}$\\
\begin{algorithmic}[1]
\STATE $tmp \leftarrow -1$, $t^* \leftarrow -1$\\
\FOR{$t=0$ to $s-1$}
\IF{$\sigma(t,x_i)=k \ \AND\  f(i-1,j-1,t)>tmp$}
\STATE $tmp \leftarrow f(i-1,j-1,t), t^* \leftarrow t$\\
\ENDIF
\ENDFOR
\RETURN $t^*$
\end{algorithmic}
\end{algorithm}

Then the value of $u$ in line 11 of Algorithm 1 must be $$u=f(i-1,j-1,t^*)=f(i-1,j-1,\max\sigma(i,j,k)).$$

We can improve the efficiency of above algorithms further in following two points.

First, we can pre-compute a table $\lambda$ of the function $\sigma(L(k)\oplus ch)$ for each character $ch\in\sum$ and $1\leq k\leq s$ to speed up the computation of $\max\sigma(i,j,k)$.
When we per-compute the prefix function $pre$, for every edge $(k,g)$ labeled with character $ch$, the value of $\lambda(k,ch)$ can be assigned directly to $g$.
The other values of the table $\lambda$ can be computed by using the prefix function $pre$ in the following recursive algorithm.

\begin{algorithm}
\caption{$\lambda(k,ch)$}
{\bf Input:} Integer $k$, character $ch$\\
{\bf Output:} Value of $\lambda(k,ch)$
\begin{algorithmic}[1]
\IF{$k>0$ \AND $\lambda(k,ch)=0$}
\STATE $\lambda(k,ch) \leftarrow \lambda(pre(k),ch)$\\
\ENDIF
\RETURN $\lambda(k,ch)$
\end{algorithmic}
\end{algorithm}

The time cost of computing all values $\lambda(k,ch)$ of the table for each character $ch\in\sum$ and $1\leq k\leq s$ by
above preprocessing algorithm is obviously $O(s|\Sigma|)$. By using this pre-computed table $\lambda$, the value of function $\sigma(L(k)\oplus ch)$ for each character $ch\in\sum$ and $1\leq k<s$ can be computed readily in $O(1)$ time.

Second, the computation of function $\max\sigma(i,j,k)$ is very time consuming and many repeated computations are overlapped in the whole {\bf for} loop of the Algorithm 1. We can amortized the computation of function $\max\sigma(i,j,k)$ to each entry of $f(i,j,k)$ in the {\bf for} loop on variable $k$ of the Algorithm 1 and finally reduce the time costs of the whole algorithm.
The modified algorithm can be described as follows.

\begin{algorithm}
\caption{M-STR-EC-LCS}
{\bf Input:} Strings $X=x_1\cdots x_n$, $Y=y_1\cdots y_m$ of lengths $n$ and $m$, respectively, and a set of $d$ constrains $P=\{P_1,\cdots,P_d\}$ of total length $r$\\
{\bf Output:} The length of an LCS of $X$ and $Y$ excluding $P$
\begin{algorithmic}[1]
\STATE Build a keyword tree $T$ for $P$
\FORALL{$i,j,k$ , $0\leq i\leq n, 0\leq j\leq m$, and $0\leq k \leq s$}
\STATE $f(i,0,k) \leftarrow 0, f(0,j,k) \leftarrow 0$ \{boundary condition\}
\ENDFOR
\FOR{$i=1$ to $n$}
\FOR{$j=1$ to $m$}
\FOR{$k=0$ to $s$}
\STATE $f(i,j,k) \leftarrow \max\{f(i-1,j,k),f(i,j-1,k)\}$
\ENDFOR
\IF {$x_i=y_j$}
\FOR{$k=0$ to $s$}
\STATE $t \leftarrow \lambda(k,x_i)$
\STATE $f(i,j,t) \leftarrow \max\{f(i,j,t),1+f(i-1,j-1,k)\}$
\ENDFOR
\ENDIF
\ENDFOR
\ENDFOR
\RETURN $\max\limits_{0\leq t<s}\{f(n,m,t)\}$
\end{algorithmic}
\end{algorithm}

Since $\lambda(k,x_i)$ can be computed in $O(1)$ time for each $x_i,1\leq i\leq n$ and any $0\leq k<s$, the loop body of above algorithm requires only $O(1)$ time. Therefore, our dynamic programming algorithm for computing the length of an LCS of $X$ and $Y$ excluding $P$ requires $O(nmr)$ time and $O(r|\Sigma|)$ preprocessing time.

Until now we have assumed that our algorithm is implemented under Assumption 1 and Assumption 2 on the constrain set $P$. We now describe how to relax the two assumptions.

If Assumption 1 is violated, then there must be some duplicated strings in the constrain set $P$. In this case, we can fist sort the strings in the constrain set $P$, then duplicated strings can be removed from $P$ easily and then Assumption 1 on the constrain set $P$ is satisfied. It is clear that removed strings will not change the solution of the problem.

For Assumption 2, we first notice that a string $A$ in the constrain set $P$ is a proper substring of string $B$ in $P$, if and only if in the keyword tree $T$ of $P$, there is a directed path of failure links from a node $v$ on the path from the root to the leaf node corresponding to string $B$ to the leaf node corresponding to string $A$ \cite{7}.
For example, in Fig.1(a), there is a directed path of failure links from node 5 to node 7 and thus we know the string $ba$ corresponding to node 7 is a proper substring of string $aba$ corresponding to node 5.

With this fact, if Assumption 2 is violated, we can remove all super-strings from the constrain set $P$ as follows.
We first build a keyword tree $T$ for the constrain set $P$, then mark all nodes passed by a directed path of failure links to a leaf node in $T$ by using a dept first traversal of $T$. All the strings corresponding to the marked leaf node can then be removed from $P$. Assumption 2 is now satisfied on the new constrain set and the keyword tree $T$ for the new constrain set is then rebuilt. It is not difficult to do this preprocessing in $O(r)$ time. It is clear that the removed super-strings will not change the solution of the problem.

If we want to get the answer LCS of $X$ and $Y$ excluding $P$, but not just its length, we can also present a simple recursive back tracing algorithm for this purpose as the following Algorithm 6.

In the end of our new algorithm, we will find an index $t$ such that $f(n,m,t)$ gives the length of an LCS of $X$ and $Y$ excluding $P$. Then, a function call $back(n,m,t)$ will produce the answer LCS accordingly.

\begin{algorithm}
\caption{$back(i,j,k)$}
{\bf Comments:} A recursive back tracing algorithm to construct the answer LCS
\begin{algorithmic}[1]
\IF{$i=0 \ \OR \ j=0$}
\RETURN
\ENDIF
\IF {$x_i=y_j$}
\IF {$f(i,j,k)=f(i-1,j-1,k)$}
\STATE $back(i-1,j-1,k)$
\ELSE
\STATE $back(i-1,j-1,\max\sigma(i,j,k))$
\PRINT $x_i$
\ENDIF
\ELSIF{$f(i-1,j,k)>f(i,j-1,k)$}
\STATE $back(i-1,j,k)$
\ELSE
\STATE $back(i,j-1,k)$
\ENDIF
\end{algorithmic}
\end{algorithm}

Since the cost of the algorithm $\max\sigma(i,j,k))$ is $O(r)$ in the worst case, the algorithm $back(i,j,k)$ will cost $O(r\max(n,m))$.

Finally we summarize our results in the following Theorem.

\begin{theo}\label{th2}
The Algorithm 5 solves the M-STR-EC-LCS problem correctly in $O(nmr)$ time and $O(nmr)$ space, with preprocessing time $O(r|\Sigma|)$.
\end{theo}

\section{Concluding Remarks}
We have suggested a new dynamic programming solution for the M-STR-EC-LCS problem.
The M-STR-IC-LCS problem is another interesting generalized constrained longest common subsequence (GC-LCS) which is very similar to the M-STR-EC-LCS problem.
The M-STR-IC-LCS problem is to find an LCS of two main sequences, in which a set of constrain strings must be included as its substrings.
It is not clear that whether the same technique of this paper can be applied to this problem to achieve an efficient algorithm. We will investigate the problem further.

\end{document}